\documentclass[preprint2]{aastex}  

\usepackage{amsmath,times,ulem}
\usepackage[displaymath]{lineno}

\shorttitle{A Critical Assessment of NLFFF Modeling}

\shortauthors{DeRosa et al.}

\def\referee#1{{#1}}

\newcommand{\sva}{SvA}

\newcommand{\bs}{\boldsymbol} 
\newcommand{\bb}{\bs B} 
\newcommand{\bj}{\bs J} 
\newcommand{\bcr}{\bs\times} 
\newcommand{\bdel}{\bs\nabla} 
\newcommand{\bdot}{\bs\cdot} 
\newcommand{\delcr}{\bdel\bcr} 
\newcommand{\deldot}{\bdel\bdot} 

\begin{document}

\renewcommand{\textfraction}{0.1}
\renewcommand{\floatpagefraction}{0.2}
\renewcommand{\dblfloatpagefraction}{0.8}
\renewcommand{\dbltopfraction}{0.8}


\title{A Critical Assessment of Nonlinear Force-Free Field Modeling of the
  Solar Corona for Active Region 10953}

\author{Marc~L.~DeRosa\altaffilmark{1}, Carolus~J.~Schrijver\altaffilmark{1},
Graham~Barnes\altaffilmark{2}, K.~D.~Leka\altaffilmark{2},
Bruce~W.~Lites\altaffilmark{3}, Markus~J.~Aschwanden\altaffilmark{1},
Tahar~Amari\altaffilmark{4,5}, Aur{\'e}lien Canou\altaffilmark{4},
James~M.~McTiernan\altaffilmark{6}, St{\'e}phane~R{\'e}gnier\altaffilmark{7},
Julia~K.~Thalmann\altaffilmark{8}, Gherardo~Valori\altaffilmark{9},
Michael~S.~Wheatland\altaffilmark{10}, Thomas~Wiegelmann\altaffilmark{8},
Mark~C.~M.~Cheung\altaffilmark{1}, Paul~A.~Conlon\altaffilmark{11},
Marcel~Fuhrmann\altaffilmark{12}, Bernd~Inhester\altaffilmark{8},
Tilaye~Tadesse\altaffilmark{8,13}}

\altaffiltext{1}{Lockheed Martin Solar and Astrophysics
Laboratory, 3251 Hanover St. B/252, Palo Alto, CA 94304, USA}

\altaffiltext{2}{NorthWest Research Associates, Colorado Research Associates
Division, 3380 Mitchell Ln., Boulder, CO 80301, USA}

\altaffiltext{3}{High Altitude Observatory, National Center for Atmospheric
Research, P.O.~Box 3000, Boulder, CO 80307, USA.  The National Center for
Atmospheric Research is sponsored by the National Science Foundation.}

\altaffiltext{4}{CNRS, Centre de Physique Th{\'e}orique de l'{\'E}cole
Polytechnique, 91128 Palaiseau Cedex, France}

\altaffiltext{5}{LESIA, Observatoire de Paris, 5 Place Jules Janssen, 92190
Meudon Cedex, France}

\altaffiltext{6}{Space Sciences Laboratory, University of California at
Berkeley, 7 Gauss Way, Berkeley, CA 94720, USA}

\altaffiltext{7}{Mathematics Institute, University of St Andrews, St Andrews,
Fife, KY16 9SS, United Kingdom}

\altaffiltext{8}{Max-Planck-Institut f{\"u}r Sonnensystemforschung,
Max-Planck-Strasse 2, 37191 Katlenburg-Lindau, Germany}

\altaffiltext{9}{Astrophysikalisches Institut Potsdam, An der Sternwarte 16,
14482 Potsdam, Germany}

\altaffiltext{10}{School of Physics, University of Sydney, Sydney, NSW 2006,
Australia}

\altaffiltext{11}{Astrophysics Research Group, School of Physics, Trinity
College Dublin, Dublin 2, Ireland}

\altaffiltext{12}{Institut f{\"u}r Physik, Universit{\"a}t Potsdam, Am Neuen
Palais 10, 14469 Potsdam, Germany}

\altaffiltext{13}{Department of Physics, Addis Ababa University, P.O.~Box
1176, Addis Ababa, Ethiopia}

\begin{abstract}

Nonlinear force-free field (NLFFF) models are thought to be viable tools for
investigating the structure, dynamics and evolution of the coronae of solar
active regions. In a series of NLFFF modeling studies, we have found that
NLFFF models are successful in application to analytic test cases, and
relatively successful when applied to numerically constructed Sun-like test
cases, but they are less successful in application to real solar
data. Different NLFFF models have been found to have markedly different field
line configurations and to provide widely varying estimates of the magnetic
free energy in the coronal volume, when applied to solar data.  NLFFF models
require consistent, force-free vector magnetic boundary data. However, vector
magnetogram observations sampling the photosphere, which is dynamic and
contains significant Lorentz and buoyancy forces, do not satisfy this
requirement, thus creating several major problems for force-free coronal
modeling efforts.  In this article, we discuss NLFFF modeling of NOAA Active
Region 10953 using Hinode/SOT-SP, Hinode/XRT, STEREO/SECCHI-EUVI, and SOHO/MDI
observations, and in the process illustrate the three such issues we judge to
be critical to the success of NLFFF modeling: (1) vector magnetic field data
covering larger areas are needed so that more electric currents associated
with the full active regions of interest are measured, (2) the modeling
algorithms need a way to accommodate the various uncertainties in the boundary
data, and (3) a more realistic physical model is needed to approximate the
photosphere-to-corona interface in order to better transform the forced
photospheric magnetograms into adequate approximations of nearly force-free
fields at the base of the corona.  We make recommendations for future modeling
efforts to overcome these as yet unsolved problems.

\end{abstract}


\section{Introduction} \label{sec:intro}

The structure and evolution of the magnetic field (and the associated electric
currents) that permeates the solar atmosphere play key roles in a variety of
dynamical processes observed to occur on the Sun.  Such processes range from
the appearance of extreme ultraviolet (EUV) and X-ray bright points, to
brightenings associated with nanoflare events, to the confinement and
redistribution of coronal loop plasma, to reconnection events, to X-ray
flares, to the onset and liftoff of the largest mass ejections.  It is
believed that many of these observed phenomena take on different morphologies
depending on the configurations of the magnetic field, and thus knowledge of
such field configurations is becoming an increasingly important factor in
discriminating between different classes of events.  The coronal topology is
thought to be a critical factor in determining, for example, why some active
regions flare, why others do not, how filaments form, and many other topics of
interest.

One model of the coronal magnetic field $\bb$ assumes that the corona is
static and free of Lorentz forces, such that $\bj\bcr\bb={\bs 0}$, where
$\bj=c\,\delcr\bb/4\pi$ is the current density.  This means that
$\delcr\bb=\alpha\bb$, and thus any electric currents must be aligned with the
magnetic field.  Because $\deldot\bb=0$, it can be shown that
$\bb\bdot\bdel\alpha=0$, demonstrating that $\alpha$ is invariant along field
lines of $\bb$.  The scalar $\alpha$ is in general a function of space and
identifies how much current flows along each field line.  In cases where
$\alpha$ varies spatially, the problem of solving for $\bb$ (and $\alpha$) is
nonlinear.  Solving for such nonlinear force-free fields (NLFFFs) requires
knowledge of $\bb$ over the complete bounding surface $S$ enclosing the
solution domain.  To be compatible with a force-free field, it is necessary
for these boundary data $\bb|_S$ to satisfy a number of consistency criteria,
which we outline in \S\ref{sec:construction} and which are explained in detail
in \citet{mol1969} and in \citet{aly1984,aly1989}.

In analyzing solar active regions, localized maps of the photospheric vector
field are typically used for the lower bounding surface $\bb|_{z_0}$, and
potential fields are used for the other surfaces.  (For the Cartesian models
discussed herein, we use the convention that the $z$ axis is normal to the
photosphere, which is located at height $z=z_0$.)  The availability of vector
field maps produced by recent instrument suites such as the Synoptic Optical
Long-term Investigations of the Sun (SOLIS) facility and the Hinode
spacecraft, building on earlier work done in Hawai`i with data from the
Haleakal{\=a} Stokes Polarimeter (HSP) and by the Imaging Vector Magnetograph
(IVM) as well as from the HAO/NSO Advanced Stokes Polarimeter (ASP) at
Sacramento Peak in New Mexico, has spurred investigations that employ
coronal-field models based on such measurements.  We anticipate that such
research will intensify when regular, space-based vector field maps from the
Helioseismic and Magnetic Imager (HMI) instrument on board the Solar Dynamics
Observatory (SDO) become available.

One goal of NLFFF modeling is to provide useful estimates of physical
quantities of interest (e.g., connectivities, free energies, and magnetic
helicities) for ensembles of active regions, so that these active regions may
be systematically analyzed and intercompared.  The use of static, force-free
models mitigates some of the computational difficulties associated with
solving the more physically realistic, time-dependent problem, as running such
dynamical models at the desired spatial and temporal resolutions for multiple
active regions typically exceeds current computing capabilities.

There exist several previous studies of individual active regions where NLFFF
models are shown to be compatible with various structures in the corona (e.g.,
\citealt{reg2002,reg2004,wie2005,reg2006,sch2008}).  Several of these studies
provide evidence of good alignment between NLFFF model field lines and the
locations of observed features such as coronal loop structures observed in EUV
and X-ray images.  Others show that the locations of sigmoids, twisted flux
ropes, and/or field line dip locations coincide with analogous features in the
NLFFF models.  Such studies are certainly encouraging, but still it remains
difficult to conclusively determine whether these models match a significant
fraction of the coronal magnetic field located within the volume overlying an
entire active region.  

As part of a long-lasting (e.g., \citealt{sak1981,mcc1997}) effort to develop
methods that generate more robust NLFFF models, a working group (in which all
of the authors of this article are participating) has held regular workshops
over the past several years.  The previous results from this collaboration are
presented in \citet{sch2006}, \citet{met2008}, and \citet{sch2008}.  Since the
launch of Hinode in 2006, we have applied multiple NLFFF modeling codes to a
few active regions for which Hinode vector magnetogram data are available and
for which nonpotential features are evident (e.g., \citealt{sch2008}).  The
resulting NLFFF models generally differ from each other in many aspects, such
as the locations and magnitudes of currents, as well as measurements of
magnetic energy in the solution domain.  In this article, we identify several
problematic issues that plague the NLFFF-modeling endeavor, and use a recent
Hinode case to illustrate these difficulties.  We describe one representative
data-preparation scheme in \S\ref{sec:construction}, followed in
\S\ref{sec:validation} by a comparison of field lines in the resulting NLFFF
models to two- and three-dimensional coronal loop paths, the latter determined
by analyzing pairs of stereoscopic images.  In \S\ref{sec:discuss}, we explain
the primary issues that we believe to impact our ability to reconstruct the
coronal field in a robust manner, and also identify and discuss the alternate
data-preparation scenarios we tried in addition to those presented in
\S\ref{sec:construction}.  Concluding remarks are presented in
\S\ref{sec:conc}.

\begin{figure*}
  \epsscale{2.0}
  \plotone{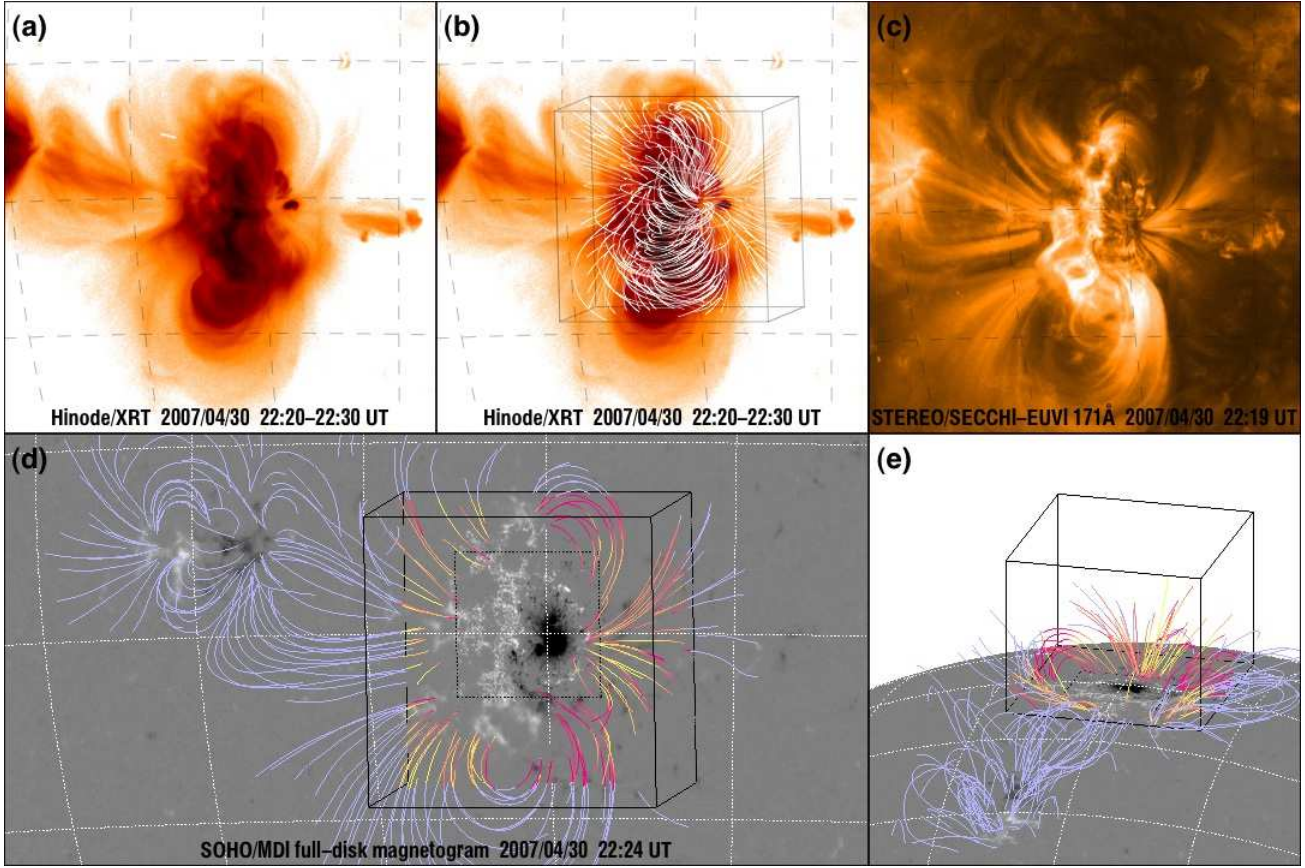}

  \caption{A series of coaligned images of AR~10953 (with the same 10$^\circ$
  gridlines drawn on all images for reference).  (a) Time-averaged and
  logarithmically scaled Hinode/XRT soft X-ray image, and (b) with the
  best-fit Wh$^-$ model field lines overlaid. (c) STEREO-A/SECCHI-EUVI 171\AA\
  image. (d) Trajectories of loops, as viewed from the perspective of an
  observer located along the Sun-Earth line of sight and determined
  stereoscopically from contemporaneous pairs of images from the two STEREO
  spacecraft.  (e) Same visualization as panel (d) but viewed from the side.
  The solid black cubes in panels (d) and (e) outline the full
  320$\times$320$\times$256-pixel NLFFF computational domain, and the interior
  dotted black square outlines the base of the smaller
  160$\times$160$\times$160-pixel volume (covering most of the Hinode/SOT-SP
  scan area) used for the field line maps of Fig.~\ref{fig3} and for the
  metrics in Table~\ref{table1}.  The STEREO-loop points are colored blue if
  outside the NLFFF computational domain, or are colored according to their
  misalignment angle $\phi$ made with the field lines from the Wh$^-$
  solution.  Yellow is indicative of $\phi<5^\circ$, red of $\phi>45^\circ$,
  with a continuous progression from yellow through orange to red for
  $5^\circ<\phi<45^\circ$.  On the bottom face of the large cube is displayed
  the $B_z$ map used during the NLFFF modeling, which includes
  higher-resolution data from Hinode/SOT-SP embedded in SOHO/MDI full-disk
  magnetogram data. The magnetogram images saturate at
  $\pm$1500~Mx~cm$^{-2}$.}

  \label{fig1}

\end{figure*}

\section{Construction of NLFFF Models for AR~10953} \label{sec:construction}

Several NLFFF extrapolation algorithms (each implementing one of the three
general classes of extrapolation methods) were applied to boundary conditions
deduced from a scan of NOAA Active Region~(AR) 10953, taken by the
Spectro-Polarimeter (SP) instrument of the Solar Optical Telescope (SOT)
\citep{tsu2008} on board the Hinode spacecraft.  The Hinode/SOT-SP scan of
this active region started at 22:30~UT on 2007 April~30 and took about 30~min
to complete.  As the scan progressed, polarization spectra of two magnetically
sensitive Fe~{\sc i} lines at 6301.5\AA\ and 6302.5\AA\ were obtained within
the 0$\farcs$16$\times$164$\arcsec$ slit, from which Stokes IQUV spectral
images were generated.  For this scan (in ``fast-map'' mode), the along-slit
and slit-scan sampling was 0$\farcs$32, and the total width of of the scan was
160$\arcsec$.  AR~10953 produced a C8.5 flare about two days after this
Hinode/SOT-SP scan, and a C4.2 flare about four and a half days after this
scan, but otherwise the active region was flare-quiet above the C1.0 level.
Images from the X-Ray Telescope (XRT) \citep{gol2007} on board Hinode around
this time show a series of bright loops in the central region of AR~10953
(Fig.~\ref{fig1}a).

The NLFFF algorithms need vector magnetic data as boundary conditions, and
determining these boundary maps comprises the first step in constructing a
NLFFF model.  The conditions pertaining to the lower boundary are determined
from a map of the photospheric vector magnetic field from the Hinode/SOT-SP
instrument.  The magnetic components parallel to and transverse to the line of
sight, $B_{\text{LOS}}$ and $\bb_t$, are functions of the circular and linear
polarization signals, respectively.  Constructing $\bb|_{z_0}$ requires
assuming an atmospheric model (in this case Milne-Eddington) and determining
which combinations of magnetic field strengths and filling factors produce the
observed polarization signals (e.g., \citealt{sku1987,kli1992,bor2007}).
$B_{\text{LOS}}$ has uncertainties that are typically an order of magnitude
less than $\bb_t$.

The next step involves removing the ambiguities in the components of $\bb_t$
that arise due to the property that the same linear polarization signal can be
produced by either of two magnetic field vectors differing by 180$^\circ$ of
azimuth in the transverse plane.  We choose to perform the disambiguation
using the interactive Azimuthal Ambiguity Method (AZAM), which is one of
several methods have been devised and tested to resolve this ambiguity (see
\citealt{met2006}, and references therein).

After disambiguation, the $\bb|_{z_0}$ map for AR~10953 is used to produce
potential field data with which the extrapolation codes will initialize the
computational domain.  Our approach is to specify the computational domain
(having an enclosing surface $S$) that contains much of the coronal volume
overlying the active region of interest, such that the lower boundary includes
the area for which vector magnetogram data are available.  The initialization
field is calculated by embedding the Hinode/SOT-SP vector magnetogram data in
a larger line-of-sight magnetogram observed by the Michelson Doppler Imager
(MDI) instrument \citep{sch1995} on board the Solar and Heliospheric
Observatory (SOHO) spacecraft (as shown in Fig. \ref{fig1}d).  Then, the
potential field coefficients corresponding to this enlarged footprint are
determined, from which the potential field in the
320$\times$320$\times$256-pixel NLFFF computational domain is computed.  In
addition, the vector field boundary conditions for the side and top boundaries
of the computational domain are taken from this same potential field
extrapolation, primarily because we expect that the coronal magnetic field
becomes largely potential away from the center of the active region, but also
because it is useful to specify how unbalanced flux emanating from this active
region connects to flux of the opposite polarity located elsewhere on the Sun.

The embedded lower-boundary data are then sampled onto a uniform, helioplanar,
320$\times$320-pixel grid having 580~km pixels, such that the footprint of the
computational domain spans a 185.6-Mm-square area.  The region for which
Hinode vector magnetogram data for AR~10953 were available comprise about a
100-Mm-by-115-Mm subarea of the full lower boundary footprint, outside of
which the horizontal components of $\bb|_{z_0}$ are set to zero.  Thus, in
this peripheral region outside the Hinode/SOT-SP field of view, the field on
the lower boundary can either be considered as purely vertical (for force-free
methods which use all three components of the field as boundary conditions),
or equivalently as having zero vertical current \referee{density} (for methods
which use the vertical component of the field together with the vertical
component of the current density).

Next, to be consistent with a force-free field, it is necessary (but not
sufficient) that the entire boundary field $\bb|_S$ satisfy several criteria,
as delineated in \citet{mol1969} and in \citet{aly1984,aly1989}: namely, (1)
the volume-integrated Lorentz force must vanish, (2) the volume-integrated
magnetic torque must vanish, and (3) the amount of negative-polarity flux
through $\bb|_S$ having a given value of $\alpha$ must equal the
positive-polarity flux through $\bb|_S$ with this same value of $\alpha$.  The
first two criteria are relations involving various components of $\bb|_S$, and
are derived from volume integrals of the Lorentz force and its first moment.
The third (``$\alpha$-correspondence'') relation operates over all values of
$\alpha$ present on $\bb|_S$.

There is of course no guarantee, however, that the values of $\bb|_{z_0}$,
coupled with the potential field of $\bb$ for the complement of the enclosing
surface, together satisfy these consistency criteria.  Our working group
attempts to deal with this problem by preprocessing the boundary data before
feeding them to the extrapolation codes.  The preprocessing scheme used here
(developed by \citealt{wie2006}) seeks to adjust the components of
$\bb|_{z_0}$ so as to satisfy the first two consistency criteria while
minimizing the deviations of $\bb|_{z_0}$ from their measured values.  During
this preprocessing step, spatial smoothing is also applied to $\bb|_{z_0}$ to
attenuate some of the small-scale magnetic fluctuations that likely die off
shortly above the photosphere.

Finally, we apply the various NLFFF algorithms to these boundary and initial
data.  Several methods for calculating NLFFF models of the coronal magnetic
field have been developed and implemented in recent years, including (1) the
optimization method, in which the solution field is evolved to minimize a
volume integral such that, if it becomes zero, the field is divergence- and
force-free \citep{whe2000,wie2004}; (2) the evolutionary magnetofrictional
method, which solves the magnetic induction equation using a velocity field
that advances the solution to a more force-free state \citep{yan1996,val2007};
and (3) Grad-Rubin-style current-field iteration procedures, in which currents
are added to the domain and the magnetic field is recomputed in an iterative
fashion \citep{gra1958,ama2006,whe2006}.  Some of these methods have been
implemented by multiple authors.  For brevity, we omit detailed explanations
of these numerical schemes as implemented here and instead direct the reader
to \citet{sch2006} and \citet{met2008}, and references therein.

Although these methods work well when applied to simple test cases
\citep{sch2006}, we have found that the results from each of the methods
typically are not consistent with each other when applied to solar data. The
resulting magnetic field configurations differ both qualitatively (e.g., in
their connectivity) as well as quantitatively (e.g., in the amount of magnetic
energy contained within them).  In discussing the results from the solar-like
test case of \citet{met2008}, we described some likely causes of such
discrepancies amongst the models.  In what follows, we illustrate these
problems in greater detail using the (solar) data set at hand.

\begin{deluxetable}{lcccc}
\tablecolumns{4}
\tablewidth{0pc}

\tablecaption{NLFFF Model Extrapolation Metrics\tablenotemark{{\it a}} for
AR~10953}

\tablehead{\colhead{Model\tablenotemark{{\it b}}} &
    \colhead{$E/E_{\text{pot}}$}\tablenotemark{{\it c}} &
    \colhead{$\left<\text{CW}\sin\theta\right>$\tablenotemark{{\it d}}$\;$} &
    \colhead{$\left<|f_i|\right>$\tablenotemark{{\it e}}($\times 10^8$)$\;$} &
    \colhead{$\left<\phi\right>$\tablenotemark{{\it f}}}}

\startdata
Pot         & 1.00 & ---  & \phantom{0}0.02 & 24$^\circ$\\
Wh$^+$      & 1.03 & 0.24 & \phantom{0}7.4\phantom{0} & 24$^\circ$\\
Tha         & 1.04 & 0.52 & 34.\phantom{00} & 25$^\circ$\\
Wh$^-$      & 1.18 & 0.16 & \phantom{0}1.9\phantom{0} & 27$^\circ$\\
Val         & 1.04 & 0.26 & 71.\phantom{00} & 28$^\circ$ \\
Am1$^-$     & 1.25 & 0.09 & \phantom{0}0.72 & 28$^\circ$\\
Am2$^-$     & 1.22 & 0.12 & \phantom{0}1.7\phantom{0} & 28$^\circ$\\
Can$^-$     & 1.24 & 0.09 & \phantom{0}1.6\phantom{0} & 28$^\circ$\\
Wie         & 1.08 & 0.46 & 20.\phantom{00} & 32$^\circ$\\
McT         & 1.15 & 0.37 & 15.\phantom{00} & 38$^\circ$\\
R{\'e}g$^-$ & 1.04\tablenotemark{{\it g}} & 0.37 & \phantom{0}6.2\phantom{0} &
42$^\circ$\\
R{\'e}g$^+$ & 0.87\tablenotemark{{\it g}} & 0.42 & \phantom{0}6.4\phantom{0} &
44$^\circ$\\
\enddata

\tablenotetext{{\it a}}{\footnotesize All metrics were evaluated over a
  160$\times$160$\times$160-pixel comparison volume (whose base overlaps much
  of the Hinode/SOT-SP scan area and is shown as a dotted line in
  Figs.~\ref{fig1}d,e), with the exception of $\left<\phi\right>$, for which
  the full 320$\times$320$\times$256-pixel computational domain was used.  The
  models are listed in order of $\left<\phi\right>$.}

\tablenotetext{{\it b}}{\footnotesize As listed in \S\ref{sec:validation}, the
  models are the initial potential solution (``Pot''); the current-field
  iteration method as run by Wheatland using the values of $\alpha$ in the
  negative (``Wh$^-$'') or positive (``Wh$^+$'') polarity; the finite-element
  Grad-Rubin-style method as run by Amari (``Am1$^-$'' and ``Am2$^-$''); the
  vector-potential Grad-Rubin-like method by Canou (``Can$^-$''), or by
  R{\'e}gnier using the values of $\alpha$ in the negative (``R{\'e}g$^-$'')
  or positive (``R{\'e}g$^+$'') polarity; the optimization method using grid
  refinement as run by Wiegelmann (``Wie'') or McTiernan (``McT''), or no grid
  refinement as run by Thalmann (``Tha''); and the magnetofrictional method
  using grid refinement  as run by Valori (``Val'').}

\tablenotetext{{\it c}}{\footnotesize $E/E_{\text{pot}}$ is the total magnetic
  energy relative to the initial potential field solution for the comparison
  volume.}

\tablenotetext{{\it d}}{\footnotesize The $\left<\text{CW}\sin\theta\right>$
  metric is the current-weighted average of $\sin\theta$, where $\theta$ is
  the angle between $\bb$ and $\bj$ in each model (with $0^\circ\le\theta\le
  180^\circ$).  For perfectly force-free fields,
  $\left<\text{CW}\sin\theta\right>=0$.}

\tablenotetext{{\it e}}{\footnotesize The $\left<|f_i|\right>$ metric is the
  mean over all pixels $i$ in the comparison volume of the absolute fractional
  flux ratio $|f_i|=|(\deldot\bb)_i|/(6|\bb|_i/\Delta x)$, where $\Delta x$ is
  the grid spacing.  The $\left<|f_i|\right>$ metric is a measure of how well
  $\deldot\bb=0$ is satisfied in the models (cf.\ eq.~[15] of
  \citealt{whe2000}), with divergence-free fields having
  $\left<|f_i|\right>=0$.}

\tablenotetext{{\it f}}{\footnotesize The quantity $\left<\phi\right>$ is the
  mean difference in angle between the stereoscopically determined loops and
  the NLFFF model field lines (with $0^\circ\le\phi\le 90^\circ$), averaged
  over the full NLFFF computational domain.}

\tablenotetext{{\it g}}{\footnotesize The R{\'e}g$^-$ and R{\'e}g$^+$
  solutions use closed boundary conditions for the side and top surfaces
  through which no magnetic flux is transmitted, and thus are associated with
  a different potential field than the Pot solution.  When comparing the
  R{\'e}g$^-$ and R{\'e}g$^+$ solutions to the potential field associated with
  these closed boundary conditions, the values of $E/E_{\text{pot}}$ are 1.23
  and 1.04, respectively.}

\label{table1}
\end{deluxetable}

\begin{figure*}
  \epsscale{1.6}
  \plotone{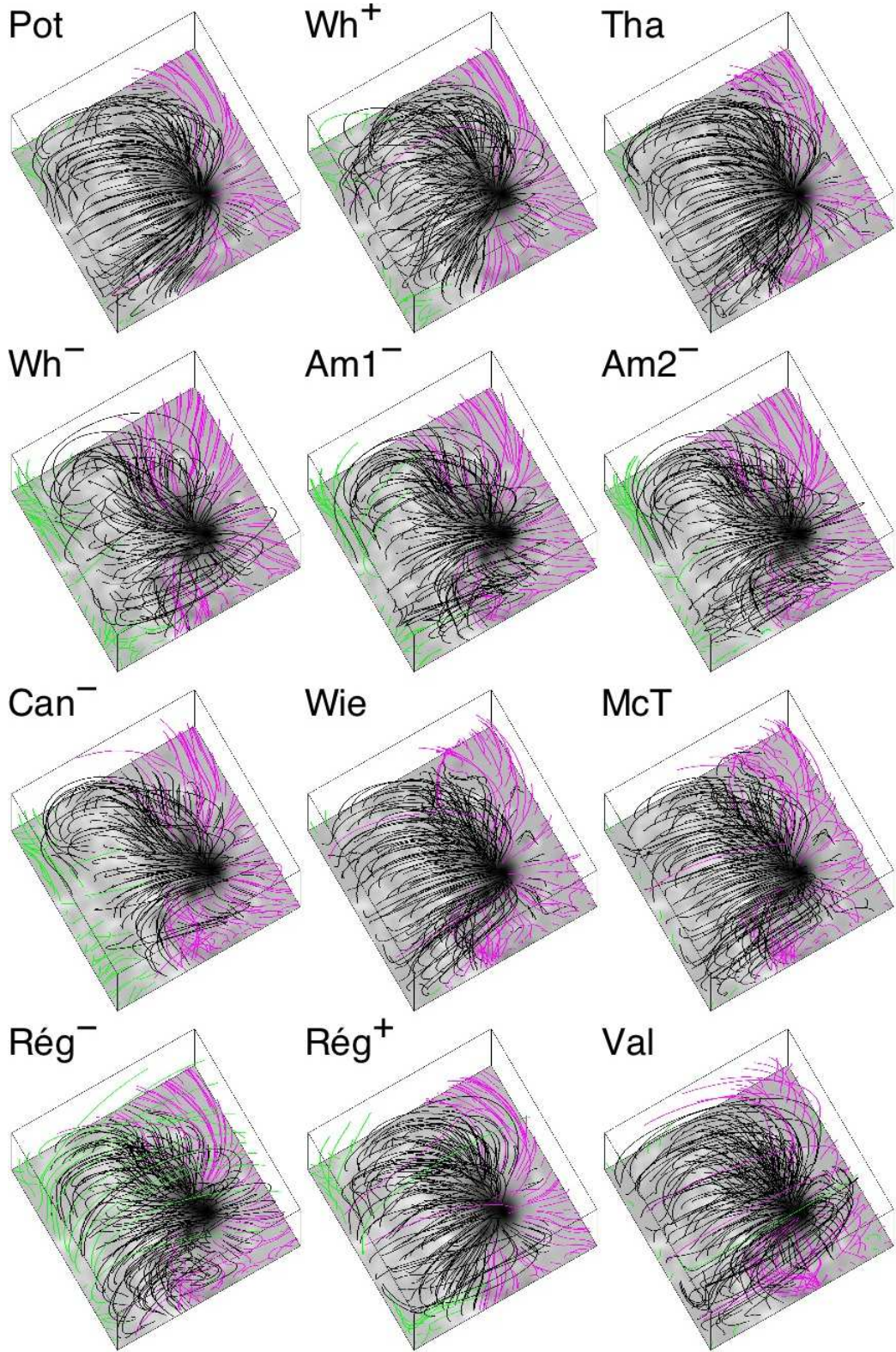}

  \caption{Representative field lines in the central portion of the active
  region for each NLFFF model listed in Table~\ref{table1}.  The cubes shown
  here comprise the same 160$\times$160$\times$160-pixel subvolumes excerpted
  from the full 320$\times$320$\times$256-pixel computational domain.  (The
  base of this subvolume is the region indicated by dotted lines in
  Figs.~\ref{fig1}d,e.)  The starting locations for the integration of the
  field lines are the same in each case, and form an array of regularly spaced
  grid points located near the lower boundary of the volume.  Black field
  lines indicate (closed) lines that intersect the lower boundary twice, and
  red and green field lines represent field lines that leave the box through
  either the sides or top, with color indicative of polarity.}

  \label{fig3}

\end{figure*}

\section{Comparison with XRT and STEREO Loops} 
\label{sec:validation}

The results of twelve extrapolations for AR~10953 (including the potential
field), based on the data-preparation steps described in
\S\ref{sec:construction}, are summarized in Table~\ref{table1} and
Figure~\ref{fig3}.  Table~\ref{table1} contains domain-averaged metrics
characterizing the center of the active region (corresponding to the region
surrounding the leading, negative-polarity sunspot), and Figure~\ref{fig3}
shows representative field lines in this same volume for each of these models.
This central region is a 160$\times$160$\times$160-pixel volume, chosen to
cover the portion of the lower boundary containing much of Hinode/SOT-SP
magnetogram data (i.e., where we have some knowledge about the currents
passing through the photosphere), and is fully contained within the larger
320$\times$320$\times$256-pixel computational domain.

The models considered in Table~\ref{table1} and Figure~\ref{fig3} are the
current-field iteration method as run by Wheatland using the values of
$\alpha$ in either the negative or positive polarity (hereafter ``Wh$^-$'' and
``Wh$^+$'', respectively); the finite-element Grad-Rubin-style method (FEMQ in
\citealt{ama2006}) run using two different parameter sets by Amari
(``Am1$^-$'' and ``Am2$^-$''); the vector-potential Grad-Rubin-like method
(XTRAPOL in \citealt{ama2006}) by Canou (``Can$^-$''), or by R{\'e}gnier using
the values of $\alpha$ in either the positive (``R{\'e}g$^+$'') or negative
(``R{\'e}g$^-$'') polarity; the optimization method using grid refinement as
run by Wiegelmann (``Wie'') or McTiernan (``McT''), or no grid refinement as
run by Thalmann (``Tha''); the magnetofrictional method using grid refinement
as run by Valori (``Val''); and the initial potential solution (``Pot'').

We find that the Am1$^-$, Am2$^-$, Can$^-$, and Wh$^-$ current-field iteration
models contain between 18\% and 25\% more energy than the potential solution,
and have smaller residual Lorentz forces and smaller average $\deldot\bb$ than
the other models.  In addition, the Am1$^-$, Am2$^-$, and Can$^-$ models find
a strongly twisted flux rope in equilibrium, whose foot points are anchored
southeast of the main spot (mostly outside of the core volume shown in
Fig.~\ref{fig3}), a feature which was anticipated by the analysis of
\citet{oka2008}.  Models using the optimization method (McT, Wie, and Tha)
contain between 4\%--15\% more energy than the potential solution, but possess
more residual Lorentz forces than the current-field iteration solutions.  The
magnetofrictional model (Val) has \referee{4\%} more energy than the
potential solution but has larger values of $\deldot\bb$ than the optimization
or current-field iteration solutions.  Based on the results summarized in
Table~\ref{table1}, the excess magnetic energy (above the potential field) for
this active region could be anywhere from near zero to about 25\% of the
potential field energy.  However, it is also possible that the excess energy
is significantly larger than 25\% when taking into account the uncertainty
associated with the inconsistency between the boundary data and the
force-free-model assumption (see \S\ref{sec:chromosphere}).

Because of these differences in the resulting NLFFF models of AR~10953, we
perform a goodness-of-fit test to determine which of the NLFFF models is the
best approximation to the observed coronal magnetic field.  In the earlier
study of \citet{sch2008}, we performed this test in both a qualitative and
quantitative manner using EUV and X-ray imagery, provided respectively by the
Transition Region and Coronal Explorer (TRACE) and Hinode/XRT instruments, by
determining which model possessed field lines that were more closely aligned
with the projected coronal loop structures visible in the (two-dimensional)
image plane.  Models for which most field lines appeared to be aligned with
loops were considered good approximations to the actual coronal magnetic
field.  Locations where the field was noticeably sheared or twisted were of
particular interest because such patterns are usually indicative of the
presence of currents (which the modeling seeks to ascertain).  More weight was
typically given to regions connected to places at the photospheric boundary
where $J_z$ is found to be high, whereas coronal loops located in the
periphery of the active region with footpoints located where $J_z$ was lower
were likely to be less sensitive to the presence of currents elsewhere in the
active region.  \referee{All such comparisons with coronal loops rest on the
  assumption that the plasma responsible for the emission is aligned with the
  coronal magnetic field and that this field is in a force-free state.}

For AR~10953, we overlaid field lines from all of the NLFFF models (as well as
the potential field model) on top of the time-averaged Hinode/XRT image shown
in Figure~\ref{fig1}a, and used the same criteria listed above to
qualitatively determine the better-matching models.  We subjectively judged
the field lines in the Wh$^-$, Am1$^-$, Am2$^-$, and Can$^-$ models to be more
closely aligned with the XRT loops than any of the others.  An overlay of
field lines from the Wh$^-$ model is shown in Figure~\ref{fig1}b.  This
judgement is based on good alignment with the tightly curved X-ray loops north
of the sunspot (which is visible in the coaligned magnetogram of this region
shown in Fig.~\ref{fig1}d), together with a reasonably good match of the loop
arcade and fan structures to the south and west of the sunspot.  This
judgement is also based on side-by-side comparisons of field line overlays
amongst the various candidate models (including the potential field model),
from which a relative ranking was determined.  The models listed above came
out on top in both instances.

With the aim of determining more quantitatively the best-fit model(s) for
AR~10953, we also compared the model field lines to three-dimensional
trajectories of loop paths.  We are able to do this because AR~10953 was
observed by the twin Solar Terrestrial Relations Observatory (STEREO)
spacecraft, one of which leads the Earth in its orbit around the Sun, and the
other of which trails the Earth.  As part of the Sun Earth Connection Coronal
and Heliospheric Investigation (SECCHI) instrument suite \citep{how2008}, each
STEREO spacecraft contains an Extreme Ultraviolet Imager (EUVI).  The angular
separation of the two STEREO spacecraft at the time AR~10953 was on disk (of
about 7$^\circ$) was favorable for stereoscopically determining the
three-dimensional trajectories of loops observed in the 171\AA, 195\AA, and
284\AA\ channels of EUVI.  The coordinates of these loop trajectories were
obtained by triangulating the positions of common features visible in pairs of
concurrent EUVI images using the method described in \citet{asc2008}.

\begin{figure}
  \epsscale{1.0}
  \plotone{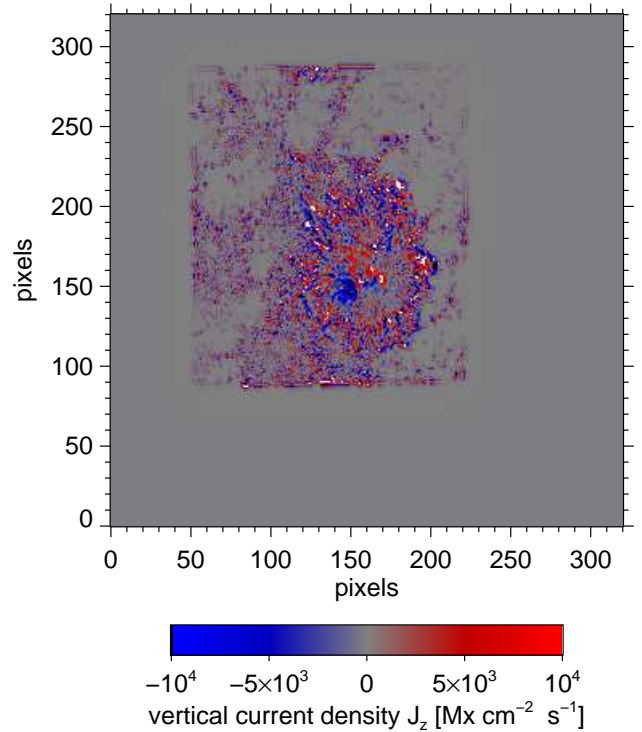}

  \caption{\referee{A map of the vertical component of the electric current
      density $J_z$ at the lower bounding surface as determined from
      Hinode/SOT-SP vector-field measurements (i.e., prior to preprocessing).
      The values of $B_x$ and $B_y$, and therefore $J_z$, outside of the
      region containing vector magnetogram data are unknown and have been
      zeroed out.  Saturation of the color table is indicated by black or
      white hues.  The pixel scale is 580~km per pixel.}}

  \label{fig6}

\end{figure}

Unfortunately, most of the loops visible in the three EUVI wavebands lie
outside of the central region of AR~10953 (Fig.~\ref{fig1}c), and thus do not
overlap the region for which the vector magnetogram data are available
(Figs.~\ref{fig1}d,e).  The main reason is that loops located closer to the
centers of the active regions tend to emit more in X-ray passbands than in EUV
passbands.  In addition, large loops at the periphery of active regions are
generally easier to reconstruct with stereoscopy, while small loops in the
centers of active regions are more difficult to discern from underlying bright
features (such as moss) and thus cannot unambiguously be triangulated.
However, the outlying loops evident in AR~10953 should still sense the
presence of currents in the center of the active region, due to Amp{\`e}re's
Law, and thus might be useful for quantitatively determining the best-matching
NLFFF model for this active region.  We infer that currents must be present in
the AR~10953 corona for two reasons.  First, most of the strong vertical
currents in the $J_z$ map are located in the central portion of the active
region \referee{(as illustrated in Fig.~\ref{fig6})} and presumably flow
upward into the corona.  Second, field lines from the potential model do not
qualitatively match the X-ray and EUV loops as well as field lines from the
Wh$^-$, Am1$^-$, Am2$^-$, and Can$^-$ models, which are our most nonpotential
models and evidently contain currents strong enough to affect the trajectories
of many field lines in the central portion of this active region
(cf.\ Fig.~\ref{fig3}).

To quantitatively compare the STEREO loops and the NLFFF-model field lines, we
determine the (positive) angle $\phi$ between the STEREO-loop and the
model-field line trajectories subtended at all STEREO-loop points lying inside
the full 320$\times$320$\times$256-pixel NLFFF computational domain.  We then
computed the mean of these angles, yielding for each model the domain-averaged
misalignment angle metric $\left<\phi\right>$ listed in Table~\ref{table1}.
We find that, at least by this particular quantitative measure, none of the
NLFFF models improve upon the value of $\left<\phi\right>=24^\circ$ found
for the potential field model, although several models (including the
qualitatively better-fitting models discussed earlier) are comparable.  We
discuss reasons why none of the models improved upon the potential field
metric for $\left<\phi\right>$ in \S\ref{sec:fovissues}.

\section{Discussion} \label{sec:discuss}

Given the boundary conditions produced using the data preparation process
described in \S\ref{sec:construction}, the various NLFFF algorithms converged
to different solutions for the coronal field above AR~10953.  A few of the
models appear to match the loop structures in the Hinode/XRT image, but none
of them were able to improve upon the potential field in their alignment with
the three-dimensional loop trajectories inferred from STEREO/SECCHI-EUVI.  In
attempting to find a consensus model, we also applied the NLFFF algorithms to
different boundary data generated using variants of the data preparation
process.  These variations, described in \S\ref{sec:dataprep}, were run in
parallel to those analyzed in \S\ref{sec:validation}, and also did not produce
a viable model.

This inability to generate models that both qualitatively and quantitatively
match the coronal loops paths is disappointing, especially given the generally
successful application of these algorithms to test cases with known solutions
\citep{sch2006}, including a solar-like test case with quasi-realistic forcing
in the lower layers that was meant to approximate some of the forces acting in
the solar chromosphere \citep{met2008}.  While we realistically expect the
various methods to yield somewhat different solutions, we cannot fully ascribe
the broad range of inconsistencies in the solutions solely to algorithmic
differences.  This causes us to examine the entire NLFFF modeling process from
beginning to end, and in so doing we have identified several additional
factors that likely also impact our ability to produce robust models.  These
factors are discussed further in \S\ref{sec:fovissues} and
\S\ref{sec:chromosphere}.

\subsection{Data Preparation Variations} \label{sec:dataprep}

We applied the NLFFF algorithms to boundary data produced using eleven
variations of the data preparation process, of which only one was outlined in
\S\ref{sec:construction}.  Variations involved substituting a different
procedure to remove the 180$^\circ$ ambiguity of the measured transverse
vector field, and/or using different versions of the standard preprocessing
algorithm.  In total, about 60 different NLFFF models for AR~10953 were
calculated.  (Not all algorithms were run on all of the available boundary
data sets.)

The first variant entailed using a different algorithm to remove the
180$^\circ$ ambiguity inherent in the vector-magnetogram inversion process.
Although there are in fact several algorithms to do this, we chose as an
alternative to AZAM to employ the automated University of Hawai`i Iterative
Method (UHIM) \citep{can1993a} because it has been used extensively in the
literature and also scored highly amongst other ambiguity resolution
algorithms \citep{met2006}.  We found that, while differences exist in, for
example, field line trajectories near regions where the ambiguity was resolved
differently, the volume-integrated metrics discussed in \S\ref{sec:validation}
and shown in Table~\ref{table1} were largely similar for both the AZAM- and
UHIM-disambiguated boundary data.

The second variant involved a new version of the method used to preprocess the
values of $\bb|_{z_0}$ to make the boundary data more consistent with a
force-free solution.  Our standard scheme pivots and smooths the components of
$\bb|_{z_0}$ so that the integrated magnetic forces and torques in the
overlying volume are reduced as much as possible, while also retaining some
fidelity to the measured vector field.  For AR~10953, we also experimented
with a preprocessing scheme (described in \citealt{wie2008}) that, in addition
to the above, seeks to align the horizontal components of $\bb|_{z_0}$ with
fibrils seen in contemporaneous images of H$\alpha$.  The motivation for this
additional preprocessing constraint is to produce boundary data as close as
possible to the force-free field expected to exist at the chromospheric level
(to which the H$\alpha$ fibrils are assumed parallel).  We found, however,
that using H$\alpha$-fibril information (observed by the Narrowband Filter
Imager of Hinode/SOT) did not make a significant difference in the
domain-averaged metrics used to characterize the various extrapolation models,
although we intend to experiment further with this preprocessing scheme as it
is somewhat new.

The third variant was to use the method of preprocessing described in
\citet{fuh2007}, the goals of which are the same as the \citet{wie2006}, but
which uses a simulated annealing numerical scheme to find the optimal
$\bb|_{z_0}$ field.  As with the other variations, using this alternate
preprocessing scheme did not much affect the resulting global metrics
\citep{fuh2009}.

\subsection{Field-of-View Issues} \label{sec:fovissues}

The Hinode/SOT-SP vector magnetogram data span only the central portion of the
AR~10953, and thus do not cover all of the weaker field and plage that
surround the active-region center.  Here, as in the \citet{sch2008} case, we
chose to extend the NLFFF computational domain and embed the vector data in a
larger line-of-sight magnetogram.  One benefit of such embedding is that it
places the side and top bounding surfaces farther away from the center of the
active region, in locations where the coronal magnetic field is presumed more
potential and thus more consistent with the boundary conditions applied there.
Another reason is that in earlier test cases using boundary data with known
solutions (described in \citealt{sch2006}), we found that enlarging the NLFFF
computational domain improved the solution field in the central region of
interest.  We attributed this behavior primarily to the sensitivity of the
final solution to the specified boundary conditions, and concluded that moving
the side and top boundaries farther away from the region of interest improved
the resulting models.

However, there is an important difference between these earlier tests and the
current case of AR~10953.  In the \citet{sch2006} study, vector data for the
entire (enlarged) lower boundary were available, and thus the locations of
currents penetrating the entire lower bounding surface, over both polarities,
were known.  In contrast, for AR~10953 we have no information about currents
located exterior to the region containing the Hinode vector magnetogram data,
\referee{as shown in Figure~\ref{fig6},}
and consequently (as stated earlier) the horizontal components of $\bb|_{z_0}$
were set to zero in the region outside of the area containing Hinode/SOT-SP
vector data.  This is obviously not correct, but lacking any knowledge of
actual horizontal fields there, this approach was presumed to be the least
damaging.  However, the lack of satisfactory results suggests that the
decision to embed may not be as harmless as originally believed.

\begin{figure*}
  \epsscale{2.0}
  \plotone{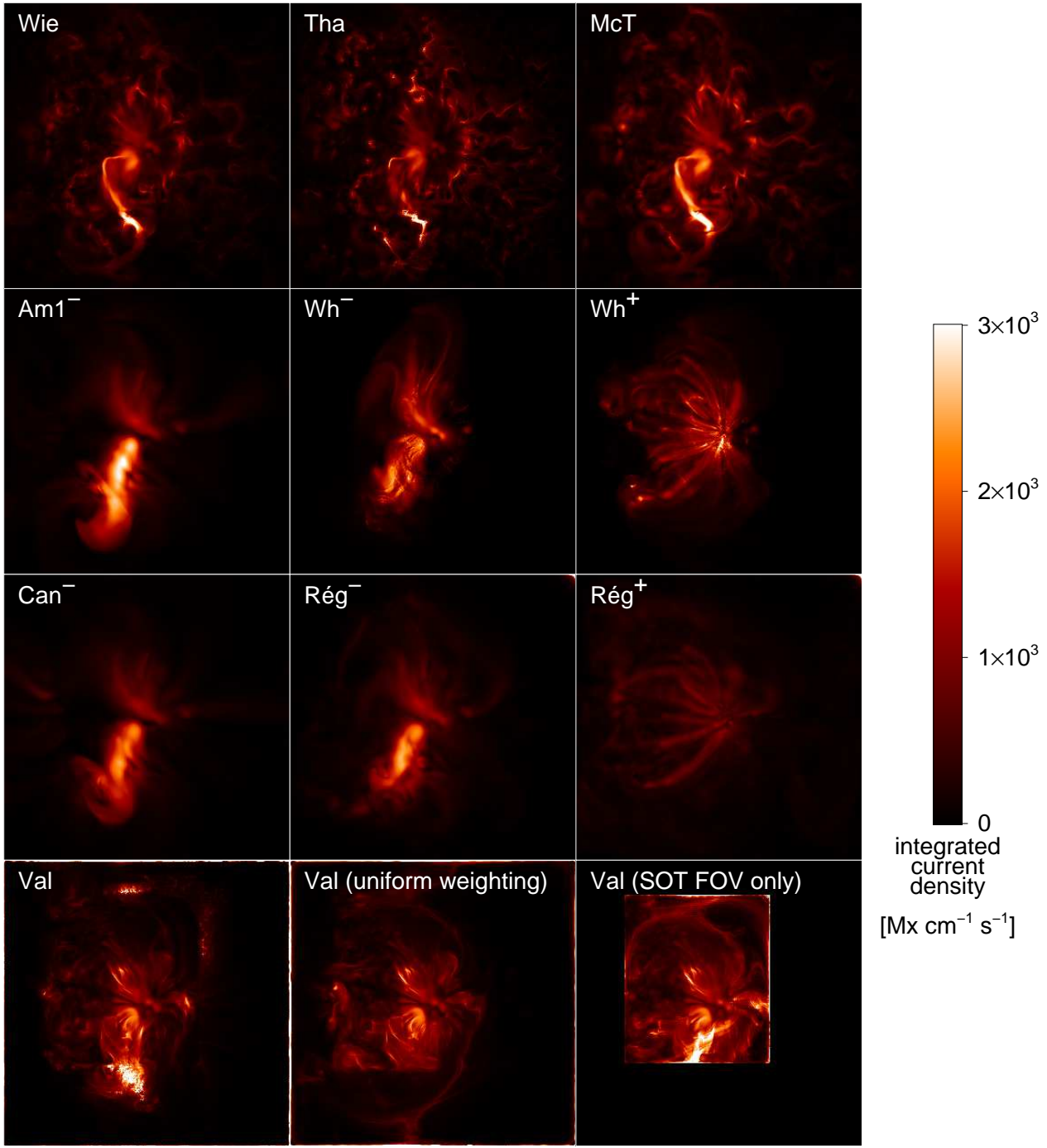}

  \caption{\referee{Images showing the magnitude of the current density
      $|\bj|$ after integrating vertically through the computational domain
      for most of the models presented in Fig.~\ref{fig3}.  Algorithms using
      the same class of method tend to produce similar patterns, as evident in
      the top row (showing models produced using optimization algorithms) and
      in the middle two rows (showing models produced using Grad-Rubin-style
      current-field iteration algorithms).  The bottom row illustrates three
      different versions of the Valori magnetofrictional model, illustrating
      some of the effects associated with the process of embedding vector
      magnetogram data into line-of-sight magnetogram data to produce
      lower-boundary data.  Shown are the Val model of Fig.~\ref{fig3} which
      weights more heavily the boundary data inside the Hinode/SOT-SP field of
      view, a model for which the lower-boundary data was weighted uniformly,
      and a smaller-domain model encapsulating only the volume overlying the
      Hinode/SOT-SP field of view.  The integrated current map from the
      Am2$^-$ model in Fig.~\ref{fig3} is almost identical to that of the
      Am1$^-$ model, and is not shown.}}

  \label{fig5}

\end{figure*}

The ability of the various NLFFF algorithms to find a valid solution
ultimately depends upon how they deal with the currents passing through the
bounding surfaces of the computational domain.  \referee{Figure~\ref{fig5}
  shows maps of the current density integrated vertically through the models.
  It is evident from these images that algorithms based on similar methods
  result in models that look similar to each other, but also that there are
  stark differences between the locations of the strong currents amongst the
  different classes of methods.}

It is interesting to note that for AR~10953, as for the \citet{sch2008} case,
the solutions bearing the best resemblance to the Hinode/XRT loops, and here
were among the best at matching the STEREO-loop trajectories, were calculated
using the current-field iteration method.  This method differs from the others
in that it uses values of $J_z$ and $\alpha$ only in one of the polarities
(the well-observed leading polarity, in the case of the best-fit models) from
the lower boundary, while ignoring such measurements in the opposite polarity.
In contrast, the optimization and magnetofrictional methods require that
information about currents be available across both polarities.

We suspect that the Wheatland current-field iteration algorithm benefits from
the additional space in the solution domain because fewer current-carrying
field lines intersect the side boundaries (which causes their values of
$\alpha$ to be set to zero).  However, the Wiegelmann optimization algorithm,
and the Valori magnetofrictional algorithm in particular, perform better when
applied to smaller volumes or when the weighting given to the peripheral
boundary information is less than that applied to the Hinode vector
magnetogram data.  \referee{The bottom row of images in Figure~\ref{fig5}
  shows that the Valori magnetofrictional algorithm has markedly different
  behavior depending on the weighting of the peripheral boundary data.  The
  differences are most striking in the area exterior to where the
  vector-magnetogram data is located.  Restricting the computational domain to
  contain only the region overlying the Hinode/SOT-SP field of view produces a
  solution with more intense currents and having fewer Lorentz forces
  ($\left<\text{CW}\sin\theta\right>=0.19$) and greater energy
  ($E/E_{\text{pot}}=1.12$) than the Val solution.}  Many of these problems
caused by the embedding process are alleviated when vector magnetogram data
are provided over a field of view that covers the locations of all relevant
currents associated with the region of interest.  For active-region studies,
this often means capturing much of the trailing polarity, which is often more
diffuse and extended than the leading polarity.

We therefore conclude that vector magnetogram data of active regions for use
by NLFFF modeling efforts need to span much of the area above which currents
flow.  Coverage of the more diffuse, trailing-polarity fields is likely to be
especially important because of the tendency for the trailing-polarity field
to contain the endpoints of many field lines that carry significant currents
(due to the existence of such currents in the leading polarity, coupled with
the assumption that many field lines connect the leading and trailing
polarities within the active region of interest).

On a related topic, we suspect that the STEREO-loop comparison process
described in \S\ref{sec:validation} is affected both by the proximity of the
STEREO loops to the sidewalls of the NLFFF computational domain (where
potential-field boundary conditions were applied) and by their lying outside
of the region for which we have vector magnetogram data (Figs.~\ref{fig1}d,e).
Consequently, one might not be surprised that the potential model bested the
others in matching the STEREO loops, but the sizable misalignment angle
$\left<\phi\right>$ of 24$^\circ$ for the potential model seems to suggest
that even these outlying STEREO loops do carry some currents.

In light of these issues, rather than using the STEREO-loop comparison as a
discriminator between the collection of NLFFF models, we instead view the
collectively poor misalignment angles by the NLFFF models as another
indication that the region over which vector magnetogram data are available
needs to be enlarged.  Although it is possible to enlarge the NLFFF
computational domain (beyond what we have already done) in order to include
even more loops observed by STEREO, we again emphasize that the added benefit
of doing so without additional vector magnetogram data would be minimal
because of the lack of further information about currents flowing through the
lower boundary.  Indeed, we applied the same current-field iteration method
used for Wh$^-$ to larger (512$\times$512-pixel) boundary data produced using
the same process described in \S\ref{sec:construction}, and found that the
value of $\left<\phi\right>$ for the identical volume used to compute the
values of $\left<\phi\right>$ in Table~\ref{table1} remained unchanged.

Lastly, we recognize that, when compared with stronger-field regions, the
transverse field components $\bb_t$ are not measurable with the same degree of
certainty in weaker-field regions such as those likely to lie within the
enlarged fields of view for which we are advocating.  The findings presented
here, however, suggest that the NLFFF modeling algorithms would benefit by
having these vector magnetic field data available, even if such data possess
higher measurement uncertainties than the stronger fields found closer to the
centers of most active regions.

\begin{figure}
  \epsscale{1.0}
  \plotone{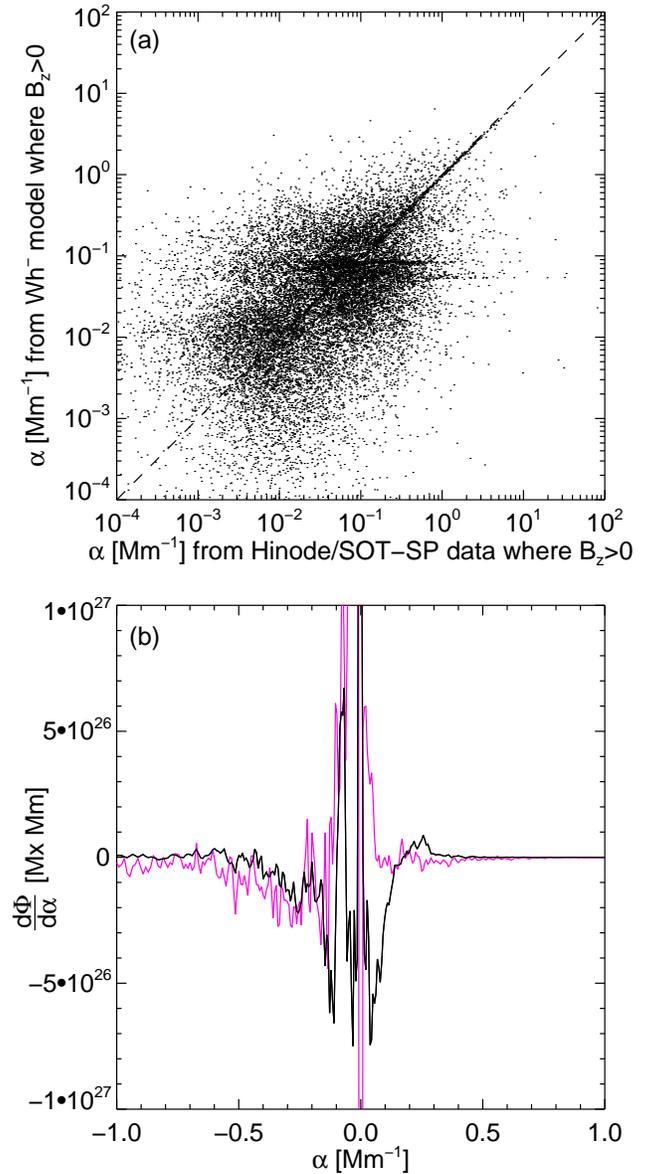}

  \caption{(a) Scatter diagram illustrating the mismatch between the values of
  $\alpha$ implied by the preprocessed Hinode/SOT-SP boundary data
  $\bb|_{z_0}$ for all points having $B_z>0$, and the values of $\alpha$ for
  field lines in the Wh$^-$ model intersecting these same points. For a
  consistent boundary condition where the $\alpha$-correspondence relation is
  satisfied, the values of $\alpha$ on each field line in the Wh$^-$ solution
  (which are taken from the negative-polarity end of the field line) would
  match the measured value of $\alpha$ found at the positive-polarity end.
  (b) The differential change $d\Phi/d\alpha$ in net flux $\Phi$ integrated
  over all points having $\alpha$ values greater than the abscissa.  The
  unpreprocessed (red) and preprocessed (black) boundary data for AR~10953 are
  both shown.  Data for which the $\alpha$-correspondence relation holds have
  $d\Phi/d\alpha=0$.}

  \label{fig2}

\end{figure}

\begin{figure}
  \epsscale{1.0}
  \plotone{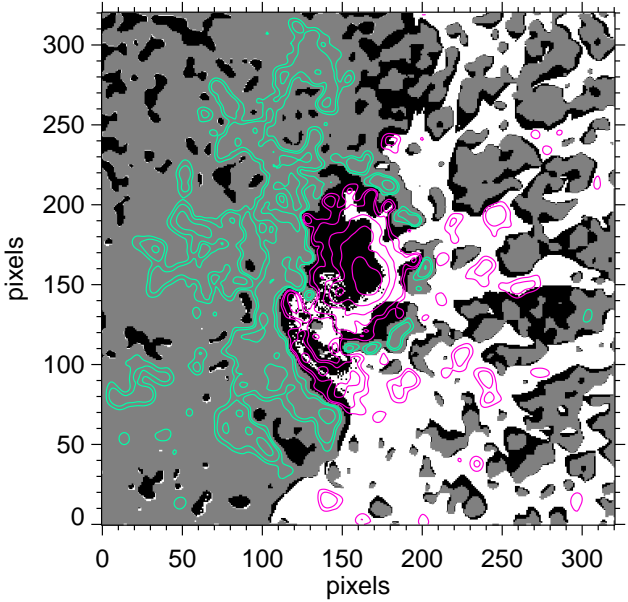}

  \caption{Censorship map for the Wh$^-$ model.  Pixels are gray in the
    positive polarity of $B_z$ and are either black or white in the negative
    polarity.  White pixels indicate the locations of field lines for which
    $\alpha$ was censored (set to zero); black pixels indicate locations where
    $\alpha\ne0$.  Contours of $B_z$ are overlaid, with green and red contours
    located in positive and negative polarity regions, respectively.  Contour
    levels are at $\pm$\{100, 200, 500, 1000, 2000\} Mx cm$^{-2}$.
    \referee{The pixel scale is 580~km per pixel.}}

  \label{fig4}

\end{figure}

\subsection{Boundary Data Inconsistencies}  \label{sec:chromosphere}

In \S\ref{sec:construction}, we described several conditions that the boundary
data $\bb|_S$ must satisfy in order to be consistent with a force-free
magnetic field.  However, these conditions are never guaranteed to be
satisfied on the full bounding surface $S$, which here consists of the vector
and line-of-sight magnetogram data for the lower boundary combined with the
potential field boundary conditions used for the remainder of the enclosing
surface.  To partially rectify this problem, we apply preprocessing to these
data to thereby adjust the various components of $\bb|_{z_0}$ on $S$ such that
the boundary data are made more compatible with the equations the NLFFF
algorithms seek to solve.

Even after preprocessing, however, the boundary data can be shown to be
incompatible with a force-free field.  The Wh$^-$ model, which is one of
several models judged to match best on a qualitative basis, only uses the
$\alpha$ values located in the negative polarity of the active region.
However, the algorithm converged to a solution for which the corresponding
$\alpha$ values in the positive polarity do not match those indicated by the
Hinode/SOT-SP data.  Figure~\ref{fig2}a illustrates this problem.  There, the
$\alpha$ values in the Wh$^-$ model from field lines that intersect the lower
boundary in the positive polarity are plotted versus the $\alpha$ values at
the same boundary points deduced from the preprocessed Hinode data.  For
consistent boundary data, these would be equal.  The scatter evident in the
figure indicates that the Hinode boundary data, even after preprocessing, are
inconsistent with a force-free field.  \referee{Additionally, the difference
  in the location of currents in the Wh$^-$ and Wh$^+$ models (and similarly
  in the R{\'e}g$^-$ and R{\'e}g$^+$ models), as evident in Figure~\ref{fig5},
  may also indicate that the boundary data are inconsistent with a force-free
  solution.}

Figure~\ref{fig2}b illustrates this effect in a different way.  This
incompatibility can be illustrated by computing
\begin{linenomath}\begin{equation}
  \Phi(\alpha) = \int_S H(\alpha'-\alpha)\, B_z\, dx\, dy,
\end{equation}\end{linenomath}
where $H$ is the Heaviside step function, and $B_z(x,y)$ and $\alpha'(x,y)$
are, respectively, the flux density and value of $J_z/B_z$ at each point on
the preprocessed Hinode boundary map.  The function $\Phi(\alpha)$ signifies
the net flux in that subarea of the boundary map for which $\alpha'$ is larger
than a certain threshold $\alpha$.  When the $\alpha$-correspondence relation
holds, the function $\Phi(\alpha)$ thus possesses a derivative of zero because
such correspondence requires, for any interval $d\alpha$, an equal amount of
positive and negative flux passing through that subarea of the boundary map
having values of $\alpha'$ between $\alpha$ and $\alpha+d\alpha$.  However,
Figure~\ref{fig2}b shows that $d\Phi/d\alpha$ is nonzero over most $\alpha$
values for the preprocessed data used here, especially within the range
$-0.2<\alpha<0.2$ which corresponds to the $\alpha$ values possessed by about
80\% of the area of the boundary map.  For comparison, the figure includes the
function $d\Phi/d\alpha$ for the unpreprocessed dataset.

The various methods deal with the lack of $\alpha$~correspondence in the
boundary data in different ways.  Current-field iteration methods allow the
$\alpha$-correspondence condition to be met by ignoring the values of $\alpha$
in one polarity. However, only limited uniqueness results have been found for
this approach, and even existence results are limited to the case of an
unbounded domain (see \citealt{ama2006}). It is well known that the
current-field iteration method fails to converge in some cases, and this may
be due to the absence of a solution, or the absence of a unique solution.  In
Wheatland's implementation of this method, if the solution does not converge,
values of $\alpha$ are censored (set to zero) in the polarity defining the
currents going into the corona. The censorship is imposed at boundary points
with $|B_z|$ less than a threshold value, and that value is increased as
required. Additional censorship is also imposed such that field lines
intersecting the side and top boundaries carry no current.  In practice it is
found that such reduction of the currents flowing into the domain can lead to
convergence.  The Wh$^-$ model, for example, censored almost half of the
values of $\alpha$ in the negative polarity (corresponding to 43\% of the
negative-polarity flux) before convergence was achieved, as illustrated in
Figure~\ref{fig4}.  Valori's magnetofrictional method is prevented from
relaxing past an equilibrium state in which the continual injection of
inconsistencies into the model (at the boundaries) is balanced by their
removal via diffusion.  Wiegelmann's optimization method does not reach as
well-relaxed of a force-free state as some of the other models, even though it
disregards some of the boundary mismatches via the tapered nature of the
weighting functions towards the edges of the model volume.

There are several reasons why the boundary conditions used for this study (and
other active region studies) might not satisfy the force-free consistency
relations.  The most conspicuous reason is that the photospheric layers of the
Sun, from which originate the Hinode/SOT-SP magnetogram data used here, do
contain Lorentz, buoyancy, and pressure gradient forces and thus are not
force-free to begin with \citep{met1995,gar2001}.  Additionally, measurement
uncertainties in the components of $\bb|_{z_0}$ preclude accurate
determinations of $J_z$ (and thus $\alpha$) on the lower boundary because of
the need to take derivatives of the horizontal components of $\bb|_{z_0}$.
Another reason is that measurements of the current \referee{density} normal to
the enclosing surface are unavailable over much of $S$ due to the lack of
vector magnetogram data above the photosphere.  Another is that the modeling
implicitly assumes that the boundary data span a planar surface, and do not
take into account effects present in vector magnetograms such as the Wilson
depression in sunspots and the broad range of line-formation heights across
the line.  Yet another is that the inversion techniques that produce the
vector magnetogram measurements do not fully take into account the multiple
components of thin, narrow strands of interleaved magnetic fields that
characterize sunspot penumbrae \citep{tit1993,bel2004,shi2008}.  We thus
conclude that the NLFFF modeling process needs to account for these intrinsic
uncertainties in the boundary data, which include everything from measurement
uncertainties to the lack of knowledge about how to infer the magnetic field
in the force-free region at the base of the corona from the observed
photospheric field maps.

\section{Conclusions} \label{sec:conc}

We have attempted to model the coronal magnetic field overlying AR~10953 by
applying a suite of NLFFF algorithms to the photospheric vector field measured
using Hinode/SOT-SP.  These data were remapped, embedded, and preprocessed in
various ways in order to produce boundary data for this active region that
were also consistent with the force-free assumption.  From these boundary
data, about 60 different NLFFF models were constructed.

The resulting variations in these models prompted us to validate the results
against images of coronal loops evident in EUV or X-ray images.  The goodness
of fit was first determined in a qualitative manner by overlaying NLFFF-model
field lines on Hinode/XRT imagery.  This comparison indicated that some models
contain field lines that are aligned with the observed loop structures.
However, conclusive determinations of best-matching models, based solely on
such overlays, remained difficult because of the indistinct nature of many
coronal loops, especially those located near the center of AR~10953 where many
of the currents are presumed to lie.

We then turned to stereoscopic determinations of three-dimensional loop paths
as a way to quantitatively assess the goodness of fit.  This comparison was
also inconclusive, because the loops traced stereoscopically in the
STEREO/SECCHI-EUVI observations were restricted to the outermost domain of the
active region.  This meant that those loops that did fall in the NLFFF
computational domain lay close to the edge of the computational volume, where
model field lines either leave the domain or run close to the side boundaries.
We suspect this quantitative comparison was at least partially compromised by
these effects, due to the model fields being sensitive to the way in which the
side boundary information is incorporated and to their being located above the
portion of the lower boundary for which Hinode/SOT-SP vector magnetogram data
were not available.

As exemplified by the qualitative and quantitative comparisons presented here,
we find that it remains difficult to construct and validate coronal magnetic
field models of solar active regions that can reliably be used for detailed
analyses of a quantitative nature.  Our experience with modeling test cases
with known solutions had shown that the various algorithms do work when given
consistent boundary conditions.  This led us to examine thoroughly the entire
NLFFF modeling framework in order to identify problematic issues that impact
our ability to build useful models of the solar coronal field.  The results of
this examination leave us with several possibilities.  First, it may be that
useful NLFFF extrapolations based on currently available signal-to-noise
levels, preprocessing procedures, fields of view, and observable fields are
intrinsically infeasible.  A second (and more hopeful) possibility is that
NLFFF extrapolations need both much larger fields of view to better constrain
the long field lines high over a region or to distant neighboring regions, and
enough spatial resolution to resolve the spatial distribution of
\referee{current densities} on the boundaries.  Third, NLFFF algorithms need
to accommodate the fact that the boundary conditions contain (sometimes
significant) uncertainties, either from the measurement process (e.g.,
signal-to-noise issues or inadequate resolution of the 180$^\circ$ ambiguity),
or from physical origins (e.g., variations in the line-formation height, or
most prominently the non-force-free nature of photospheric vector
magnetograms).

The second possibility can be tested empirically.  One way to do this with
current codes and instrumentation is to obtain vector magnetic observations of
a substantially smaller active region and its wide surroundings.  This will
place the side boundaries relatively farther away from the region of interest,
while remaining compatible with the range and resolution of, e.g., the
Hinode/SOT-SP and with the Cartesian nature of the available modeling codes.

To address the third possibility, we have several avenues available.  Simple
ways to account for boundary data uncertainties include introducing a
position-dependent weighting function used in relaxation methods, or modifying
the selection criteria for the $\alpha$ field in the current-field iterative
method.  Additionally, the preprocessing of the raw vector data needs to
better approximate the physics of the photosphere-to-chromosphere interface in
order to transform the observed photospheric field to a realistic
approximation of the overlying near-force-free field at the base of the
corona.  One way to do that without resorting to more computationally
intensive MHD models is to use the magnetohydrostatic concept (e.g.,
\citealt{wie2006a}) and approximate the stratifications for the flux tubes and
their surroundings (or the strongly and weakly magnetic regions) separately.

Finally, in light of our findings in this study and in consideration of the
aforementioned goal of constructing models that provide useful estimates of
physical quantities of interest, we thus recommend that a particular
force-free extrapolation should not be considered a consistent model of an
active-region corona unless the following indicators (at a minimum) are
satisfied: (1) good alignment of modeled field lines to the coronal loops
observed on the solar disk; (2) acceptable agreement of the
$\alpha$-correspondence relation by having similar values of $\alpha$ at both
ends of all closed field lines, and acceptable agreement with the boundary
values of $\alpha$ from the data; while (3) still realizing low values of the
NLFFF metrics $\left<\text{CW}\sin\theta\right>$ and $\left<|f_i|\right>$.

\acknowledgements

We gratefully acknowledge Prof.\ Dr.\ Sami Solanki and the Max-Planck-Institut
f{\"ur} Sonnensystemforschung in Katlenburg-Lindau, Germany, for their
hospitality during our most recent workshop, at which the ideas presented in
this article were discussed and refined.  Hinode is a Japanese mission
developed and launched by ISAS/JAXA (Japan), with NAOJ as domestic partner and
NASA (USA) and STFC (UK) as international partners. It is operated by these
agencies in cooperation with ESA and NSC (Norway).  The STEREO/SECCHI data
used here are produced by an international consortium of the Naval Research
Laboratory (USA), Lockheed Martin Solar and Astrophysics Laboratory (USA),
NASA/Goddard Space Flight Center (USA), Rutherford Appleton Laboratory (UK),
University of Birmingham (UK), Max-Planck-Institut f{\"ur}
Sonnensystemforschung (Germany), Centre Spatiale de Li{\`e}ge (Belgium),
Institut d'Optique Th{\'e}orique et Appliqu{\'e}e (France), and Institut
d'Astrophysique Spatiale (France).  M.L.D., C.J.S., G.B., and K.D.L. were
supported by Lockheed Martin Independent Research Funds.  \referee{M.L.D. was
  also supported by NASA contract NNM07AA01C to Lockheed Martin.}  J.M.M. was
supported by NASA grants NNG05144G and NNX08A156G.  S.R. acknowledges the
financial support of the UK STFC.  J.K.T. acknowledges support from DFG grant
WI~3211/1-1.  G.V. was supported by DFG grant HO~1424/9-1.  T.W. acknowledges
support from DLR grant 50~OC~0501.  P.A.C. is an IRCSET Government of Ireland
Scholar.  T.T. acknowledges support from the International Max Planck Research
School on Physical Processes in the Solar System and Beyond.

\facility{{\it Facilities}: Hinode, SOHO, STEREO}

\end{document}